\documentclass[12pt]{article}
\usepackage{amssymb}
\usepackage{amsmath}
\usepackage[dvipdfmx]{graphicx}
\usepackage{cite}
\begin{document}

\begin{titlepage}

\begin{flushright}
ICRR-Report 673-2013-22
\end{flushright}

\vskip 1.35cm

\begin{center}

{\large 
{\bf Neutrino masses, leptogenesis, and  sterile neutrino dark matter} 
}

\vskip 1.2cm
\renewcommand\thefootnote{*}
Takanao Tsuyuki\footnote{E-mail: tsuyuki@muse.sc.niigata-u.ac.jp}\\
\renewcommand\thefootnote{\arabic{footnote}}
\setcounter{footnote}{0}

\vskip 0.4cm

{ \it Graduate School of Science and Technology, Niigata University, Niigata 950-2181, Japan}
{ \it Institute for Cosmic Ray Research,
University of Tokyo, Kashiwa 277-8582, Japan}

\date{\today}

\begin{abstract} 
We analyze a scenario in which the lightest heavy neutrino $N_1$ is a dark matter candidate and the second-heaviest neutrino $N_2$ decays producing lepton number. If $N_1$ were in thermal equilibrium, its energy density today would be much larger than that of the observed dark matter, so we consider energy injection by the decay of $N_2$. In this paper, we show the parameters of this scenario that give the correct abundances of dark matter and baryonic matter and also induce the observed neutrino masses. This model can explain a possible sterile neutrino dark matter signal of $M_1$=7 keV in the x-ray observation of x-ray multi-mirror mission.
\end{abstract}

\end{center}
\end{titlepage}

\section{Introduction}
There are at least three phenomena that cannot be explained within the standard model of particle physics (SM). They are neutrino masses, dark matter (DM), and baryon asymmetry of the Universe (BAU). In the SM, particles acquire masses by the Higgs mechanism, but neutrinos are assumed not to couple to the Higgs particle so they remain massless. By introducing right-handed neutrinos, neutrinos can couple to the Higgs particle, and acquire Dirac masses. Right-handed neutrinos are singlets under the SM gauge group $SU(2)_L\times U(1)_Y$, so they can also have Majorana masses without breaking the symmetry of the SM. If these Majorana masses are much larger than the Dirac masses, mass eigenstates are separated into two groups. One of them is active, light neutrinos mainly composed of left-handed neutrinos, and the other is sterile, heavy neutrinos almost coinciding with right-handed neutrinos. This is called the seesaw mechanism \cite{Minkowski:1977sc,Yanagida:1979as,gell1979r}, and we use this mechanism in this paper.

Two other phenomena beyond the SM can be found in the Universe. In the observable range of the Universe, no primordial antimatter has been found. To explain the asymmetry of matter and antimatter, we need $CP$ violation \cite{Sakharov:1967dj}. By observation of the anisotropy of cosmic microwave background \cite{Ade:2013ktc}, the ratio of baryon density to critical density today is $\Omega_b h^2= 0.02214\pm 0.00024$(68\% limit), which can be converted into the baryon number to entropy ratio $Y_B\simeq 0.86\times 10^{-10}$. In the SM, $CP$ is violated by Yukawa couplings between the quarks and Higgs particle, but it is too small to explain this asymmetry. If there are right-handed neutrinos, $CP$ can also be broken by the couplings between the neutrinos and Higgs particle. A heavy neutrino, which is Majorana particle, it can decay into either lepton or antilepton by Yukawa interaction. The difference between these decay rates results in nonzero lepton number (called leptogenesis \cite{Fukugita:1986hr}), and it is transferred to a baryon number by electroweak processes (sphaleron processes \cite{Kuzmin:1985mm}). 

Another problem the Universe offers us is dark matter. Dark matter does not interact with electromagnetic forces and is stable, or its lifetime is longer than the age of the Universe. In the SM, such a particle is not included, and even massive left-handed neutrinos are too light to explain all dark matter. Many candidates of dark matter have been proposed, and sterile neutrino is one of them. Sterile neutrinos, of course, do not participate in electromagnetic nor strong interactions. If their mass is roughly of order keV, they can live longer than the Universe, so they can be dark matter (see Refs. \cite{Kusenko:2009up,Merle:2013gea} for reviews). There are several ways to produce sterile neutrino dark matter $N_1$. The simplest production mechanism is to use the mixing of sterile and active neutrinos, proposed Dodelson and Widrow (DW mechanism, \cite{Dodelson:1993je}). This model, however, is disfavored by observations of x-rays and Lyman alpha forest \cite{Asaka:2006nq}. Another way is resonant production, or the Shi-Fuller mechanism \cite{Shi:1998km}. If there were relatively large lepton number (at least $Y_L\gtrsim 8\times10^{-6} $), light neutrinos can be efficiently converted to sterile neutrinos. In this case, some mechanism is needed to make this lepton number much larger than the baryon number after the freeze-out of sphalerons. In the neutrino minimal standard model ($\nu$MSM), which is an extension of the SM with right-handed neutrinos with masses smaller than the electroweak scale\cite{Asaka:2005an,Asaka:2005pn}, this lepton asymmetry is produced by the decay of sterile neutrinos $N_2,\; N_3$ with masses $\gtrsim 100$MeV. In this model, their masses need to be highly degenerate, roughly $(M_3-M_2)/M_2\lesssim 10^{-3}$  to produce BAU \cite{Laine:2008pg,Canetti:2012kh}. 
Other production mechanisms need more extensions of the SM, such as decays of scalar fields \cite{Petraki:2007gq,Merle:2013wta} or new gauge interactions. We consider the last case in this paper. The Majorana mass term of the right-handed neutrinos appears as the result of a gauge symmetry breaking, and its scale is much higher than the electroweak scale. This naturally happens if grand unification exists at high energy.

In this paper, we suppose the lightest sterile neutrino $N_1$ constitutes all dark matter and the second-lightest one $N_2$ causes leptogenesis, and they were in thermal equilibrium by a gauge interaction of right-handed neutrinos \cite{Bezrukov:2009th}. There are many advantages in this case. We need not assume an initial abundance of $N_1$ and $N_2$. Their abundance is completely determined by statistical mechanics without uncertainty. The lepton number is efficiently produced, because there is no cancelation of lepton asymmetry which happens if $N_2$ is produced by Yukawa interaction. The temperature of $N_1$ is colder compared to the DW mechanism case, so constraint from Lyman alpha forest is weakened. 

The drawback of thermal relic $N_1$ is its overproduction. This problem can be solved by the decay of an out-of-equilibrium particle. Such a decay gives energy into the thermal bath, and the temperature of the thermal bath drops slowly compared to that of the decoupled particle $N_1$. The energy ratio of $N_1$ today becomes smaller, so the problem of overproduction can be solved. Cases of low-scale new gauge interaction were considered in Refs. \cite{Bezrukov:2009th,Nemevsek:2012cd}. 
We consider a case that the dark matter $N_1$ was diluted by the entropy production during the leptogenesis. This idea was proposed in Ref. \cite{Bezrukov:2012as}. They estimated the orders of $M_2, M_3$ and scale of gauge interaction of right-handed neutrinos. We refine their analysis, considering seesaw mechanism and various constraints on parameters more seriously. We explicitly show the parameters which can explain observed neutrino masses, the BAU,  and dark matter abundance. 

As a result, we found that a Majorana mass term of left-handed neutrinos $M_L$ is essential for masses of active neutrinos $m_\nu$, since the two eigenvalues of the difference $X_\nu\equiv m_\nu-M_L$ need to be very small, $X_1\lesssim 10^{-10}{\rm eV},\;X_2\lesssim 10^{-5}$eV. The third eigenvalue is much larger, $X_3\gtrsim O(0.1)$eV, in order to produce the BAU. If there is no fine-tuning, $M_2\gtrsim O(10^{8})$GeV and the scale of gauge interaction of right-handed neutrinos is $G_{FR}^{-1/2}\gtrsim 10^{12}$GeV. Recently, an unidentified line at 3.5keV in x-ray spectra was found \cite{Bulbul:2014sua,Boyarsky:2014jta}. There are many works to explain this anomaly (see, for example, Refs. \cite{Frandsen:2014lfa,Abazajian:2014gza} If this photon was emitted by dark matter, our model can explain it by decay of $N_1$, with $M_1$=7keV and $X_2+X_3|R_{31}|^2\sim 1\times 10^{-7}$eV ($R_{31}$ is a parameter of Yukawa coupling).

We use the left-right symmetric model \cite{Mohapatra:1974hk} as an example, but if $N_1$ and $N_2$ can be in thermal equilibrium, any other model is possible. Our discussion does not involve the detail of the new gauge interaction that we will introduce. Note that the idea of diluting dark matter by leptogenesis can be applied to other particles that freeze out before the decay of the heavy neutrino. 

This paper is organized as follows. In Secs. \ref{numass}-\ref{BAU}, we describe how right-handed neutrinos can explain three beyond the SM phenomena described above. In Sec. \ref{COP}, we summarize constraints on parameters from various observations and our thermal history scenario. In Sec. \ref{Ana} we show parameters that satisfy all conditions obtained in previous sections, and compare the result with observations.

\label{sec:intro}

\section{Neutrino masses} \label{numass}
We assume three right-handed neutrinos $\nu_R$ exist. The most general mass term of neutrinos can be written as
\begin{align}
\mathcal {-L}_{mass} 
 &= \frac{1}{2}\begin{pmatrix} \overline{\nu_L} & \overline{\nu^c_R}  \end{pmatrix} \begin{pmatrix} M_L& m_D \\ m_D^T  & M_R \end{pmatrix}\begin{pmatrix} \nu^c_L \\ \nu_R \end{pmatrix} +\rm h.c., \label{Bez2}
\end{align}
where $M_L$ and  $M_R$ denote $3\times 3$ Majorana mass matrices of left and right neutrinos and $m_D$ represents $3\times 3$ Dirac mass matrix. For example, this term appears in a model that has the symmetries of $SU(2)_L\times SU(2)_R\times U(1)_{B-L}$ (called the left-right symmetric model \cite{Mohapatra:1974hk,Mohapatra:1974gc,Mohapatra:1980yp}) as 
\begin{equation}
-\mathcal L_{mass}=h_{\alpha\beta} \overline{L_L^\alpha} \phi L_R^\beta+g_{\alpha\beta } \overline{ L_L^\alpha}  \tilde{ \phi} L_R^\beta+\frac{f_{\alpha\beta }}{2}(\overline{L_L^{\alpha c}} i\tau_2 \Delta_L L_L^\beta+\overline{L_{R}^{\alpha c}} i\tau_2 \Delta_R L_R^\beta)+\rm h.c.,
\end{equation}
 where $L_{L,R}$ represent SU(2)$_{L,R}$ doublets of left- or right-handed leptons, 
 \begin{align}
 L_{L,R\alpha}\equiv \begin{pmatrix} \nu_{L,R\alpha}\\ l_{L,R\alpha}\end{pmatrix},\;\alpha=e,\mu,\tau ,
  \end{align}
and $\phi, \Delta_{L,R}$ are the $SU(2)_L\times SU(2)_R$ Higgs bidoublet and triplets, which acquire vacuum expectation values
\begin{equation} 
  \langle \Delta_{L,R} \rangle=\begin{pmatrix} 0& 0 \\ v_{L,R}& 0  \end{pmatrix},\;
    \langle \phi \rangle=\begin{pmatrix} v_1& 0 \\ 0& v_2  \end{pmatrix}, \; 
    |v_1|^2+|v_2|^2=v^2=(174\rm GeV)^2,
\end{equation}
and $\tilde{\phi}\equiv \tau_2\phi^*\tau_2$. By defining
\begin{align} 
 m_{D\alpha\beta}\equiv y_{\alpha\beta }v &\equiv h_{\alpha\beta }v_1+g_{\alpha\beta }v^*_2,\quad M_{L\; \alpha\beta }\equiv f_{\alpha\beta }^*v_{L}^*,\;M_{R\; \alpha\beta }\equiv f_{\alpha\beta }v_{R},
  \end{align}
we recover Eq. (\ref{Bez2}).

Assuming the orders of $M_L$ and $m_D=yv$ are much smaller than that of $M_R$, we get mass eigenvalues,
 \begin{align}
  \mathcal {-L}_{mass}  &= \frac{1}{2}\begin{pmatrix} \overline{\nu'_L} & \overline{\nu'^c_R}  \end{pmatrix} \begin{pmatrix} m_{\nu}^{\rm {diag}}& 0 \\ 0  & M_N^{\rm {diag}} \end{pmatrix}\begin{pmatrix} \nu'^c_L \\ \nu'_R \end{pmatrix} +\rm h.c. ,\\
m_{\nu}^{\rm {diag}} &= U^{\dag} m_\nu U^*={\rm diag}[m_1, m_2, m_3], \label{Bezrukov57}\\
M_N^{\rm{diag}} &= V^{T} M_N V={\rm diag}[M_1, M_2, M_3],  \\
m_\nu &= M_L-v^2yM_R^{-1}y^{T}, \label{Bezrukov46}\\
M_N &= M_R.
\end{align}
Equation (\ref{Bezrukov46}) is the seesaw relation \cite{Magg:1980ut}, which gives a relation between masses of light neutrinos and heavy neutrinos. The matrix $U$, which diagonalizes $m_\nu$, is the Pontecorvo-Maki-Nakagawa-Sakata matrix. Flavor eigenstates $\nu_{L\alpha}$ can be written with mass eigenstates $\nu_{Li}',\; \nu_{RI}'^c (i,I=1,2,3)$ as
  \begin{align}
 \nu_{L\alpha}&= U_{\alpha i}\nu'_{Li}+\Theta_{\alpha I}\nu'^c_{RI} \label{D18},\\
 \Theta_{\alpha I}&\equiv (m_D M_R^{-1} V^*)_{\alpha I}.
\end{align}
The strength of interaction of $\nu_{RI}$ through this mixing is parametrized by the active-sterile mixing angle,
\begin{align}
\Theta_I^2\equiv \sum_{\alpha}|\Theta_{\alpha I}|^2.
 \end{align}
The flavor eigenstates of right-handed neutrinos are, ignoring $O(\Theta)$ terms,
\begin{align}
\nu_{R\alpha}\simeq V_{\alpha I}\nu_{R I}'
 \end{align}
It is conventional to define Majorana fields $N_I\equiv \nu_{RI}'+\nu_{RI}'^c$ . They are called sterile or heavy neutrinos. $I=1,2,3$ are ordered to give $M_1<M_2<M_3$.

\section{Dark matter and entropy production}

We suppose dark matter consists entirely of $N_1 \;(M_1\sim O(1)\rm keV)$ and that $N_2 \;(M_2\gtrsim 10^{9}\rm GeV)$ causes leptogenesis. The thermal history we consider in this paper is as follows \cite{Bezrukov:2012as}. $N_1$ and $N_2$ were once in thermal equilibrium, and they froze out when they were still relativistic ($T_f\gtrsim 10^{10}$GeV). As the temperature went down, $N_2$ became nonrelativistic and dominated the Universe. $N_2$ decays at $T\gtrsim 10^{5}$GeV, which is much higher than sphaleron freeze-out temperature. This decay created lepton number and gave energy to the thermal bath, i.e., entropy was produced. This lepton number was transformed into the BAU observed today. By this entropy production, the abundance of $N_1$ became the observed DM abundance, $\Omega_{DM} h^2\simeq 0.12$.

The Yukawa interaction of $N_1$ is too weak to bring it to thermal equilibrium. We have to introduce a new gauge interaction between right-handed neutrinos and the SM particles \cite{Bezrukov:2009th}. In the left-right symmetric model, the gauge interaction term of right-handed neutrinos is written as \cite{Zhang:2007da}
\begin{align}
-\mathcal{L}_{R,int}=g_R \overline{L_{R}}W_R^\mu\gamma_\mu L_R.
 \end{align}
Right-handed neutrinos freeze out from thermal bath when the rate of gauge interaction\footnote{We assume the scale of  $v_R$ is larger than $T_f$.}  equals to the Hubble rate,
\begin{equation}
G_{FR}^2T_f^5\sim \sqrt{g_{*f}}\frac{T_f^2}{M_{Pl}} ,
    \label{G_N}
\end{equation}
where $M_{Pl}=1.22\times~10^{19}$GeV, $G_{FR}\equiv \frac{\sqrt{2}g_R^2}{8M'^2}$ and $M'$ represents the mass of the new gauge boson. For the effective degrees of freedom of particles in the thermal bath, we take $g_{*f}\sim 110$. By solving Eq. (\ref{G_N}), we obtain the freeze-out temperature
 \begin{align} 
 T_f &\sim  10^{10} {\rm GeV}\left(\frac{(10^{12}\rm GeV)^{-2}}{G_{FR}}\right)^{2/3}. \label{Tf}
 \end{align}
We assume heavy neutrinos $N_1$ and $N_2$ decouple when they are still relativistic, i.e., $T_f>M_2$. Then the yield of $N_1$ after freeze out is
\begin{align}
 \left.Y_{N_1}\right|_f\equiv\left.\frac{n_{N_1}}{s}\right|_f=\frac{1}{g_{*f}}\frac{135\zeta(3)}{4\pi^4} = \frac{0.416}{g_{*f}} \label{Y}.
\end{align}
This is constant if there is no entropy production. The energy ratio of $N_1$ today is
 \begin{align}
   \Omega_{N_1} &=  9.5\Omega_{DM}\frac{110}{g_{*f}}\frac{M_1}{\rm keV}\frac{S_f}{S_0} ,\label{ON1}
\end{align} 
where $S_f$ and $S_0$ represent entropy in comoving volume at $N_1$ freeze-out and today. From the observation of dwarf spheroidal galaxies, the mass of the fermion dark matter particle must be larger than 1 keV \cite{Gorbunov:2008ka}. If $\Omega_{N_1}=\Omega_{DM}=0.12/h^2$ \cite{Ade:2013ktc}, entropy needs to become about 10 times larger. 

\begin{figure}[t]
  \centering
 \includegraphics{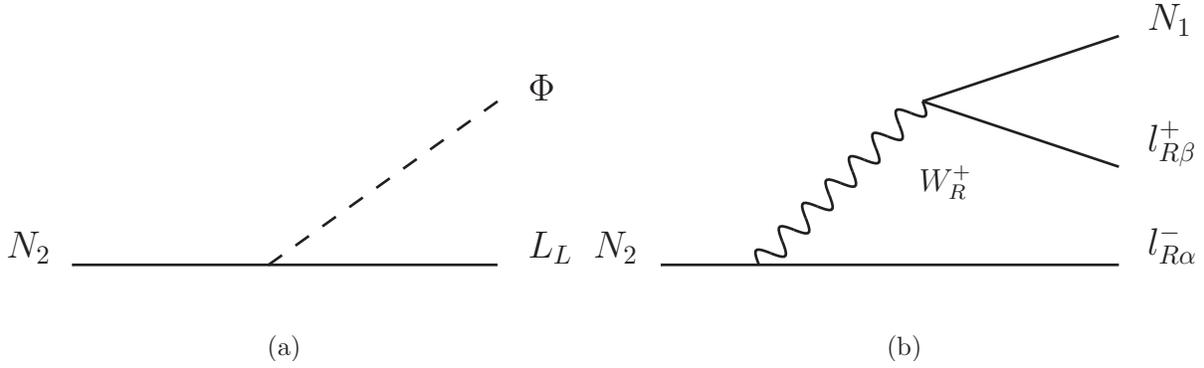}
  \caption{Decay of $N_2$ by Yukawa interaction (a) and gauge interaction (b).
  }  \label{NllN}
\end{figure}

We consider entropy production by the decay of $N_2$. The ratio of $S_0$ to $S_f$ can be expressed by $\Gamma_{N_2}$, the decay rate of $N_2$, as
\begin{align} 
\frac{S_0}{S_f} &= 0.76\frac{\langle g_*^{1/3}\rangle^{3/4} M_2}{g_{*f}\sqrt{\Gamma_{N_2} M_{Pl}}},
\end{align}
where $\langle g_*^{1/3}\rangle$ is the average of $g_*^{1/3}$ during the decay \cite{Scherrer:1984fd}. Substituting this into Eq. (\ref{ON1}), and from the condition $\Omega_{N_1}=\Omega_{DM}$, the decay rate must satisfy
\begin{align}
\Gamma_{N_2}  &= 0.50 \times 10^{-6}\langle g_*^{1/3}\rangle^{3/2}  \frac{M_2^2}{M_{Pl}}\left( \frac{{\rm keV}}{M_1} \right)^2.
\label{GN2}
 \end{align}
 There are two decay channels of $N_2$ at tree level (see Fig. \ref{NllN}). To generate lepton number successfully, the decays by Yukawa interaction have to be dominant, 
 \begin{align}
\Gamma_{N_2}\simeq\Gamma_{N_2\to L_L \Phi}&=\frac{(\tilde{y}^\dag \tilde{y})_{22}}{8\pi}M_2,
 \end{align}
where $\tilde{y}\equiv yV$. We will check the condition for this approximation in Sec. \ref{COP}. Then,  we obtain a condition for Yukawa coupling constants,
\begin{align}
(\tilde{y}^\dag \tilde{y})_{22} = 1.1\times 10^{-14} \left( \frac{{\rm keV}}{M_1} \right)^2 \left( \frac{M_2}{10^9\rm GeV} \right).
\label{yy22}
 \end{align}
 
 \section{Baryon asymmetry} \label{BAU}
$N_2$ is a Majorana field, so it can decay into both lepton and antilepton. The difference between these decay rates results in a net lepton number production \cite{Fukugita:1986hr}. This lepton number is transformed into baryon number by electroweak processes \cite{Kuzmin:1985mm}. By integrating the Boltzmann equations for $N_2$ and leptons, we obtain the yield of the baryon number today:
\begin{align}
Y_B &=-\frac{28}{79} \frac{0.416}{g_{*f}}\epsilon\frac{S_f}{S_0}= -1.4\times 10^{-4}\epsilon \left( \frac{ {\rm keV}}{M_1} \right).
 \end{align}
 $\epsilon$ denotes the asymmetry of decay rates, calculated as \cite{Covi:1996wh}
 \begin{align} 
\epsilon&= -\frac{1}{8\pi}\frac{{\rm Im}[((\tilde{y}^\dag \tilde{y})_{32})^2]}{(\tilde{y}^\dag \tilde{y})_{22}}g(M_3^2/M_2^2),\label{epsi}\\
g(x)&\equiv\sqrt{x}\left(\frac{1}{1-x}+1-(1+x)\ln\frac{1+x}{x}\right). \label{Fong2.30}
 \end{align}
We ignored the contribution of $N_1$, since $(\tilde{y}^\dag \tilde{y})_{12}$ is much smaller than $(\tilde{y}^\dag \tilde{y})_{32}$ in our scenario (see Sec. \ref{Ana}). By the observation of cosmic microwave background \cite{Ade:2013ktc}, $Y_B\simeq0.86\times 10^{-10}$, so 
\begin{align}
\epsilon = -6.1 \times 10^{-7} \frac{M_1}{\rm keV}
\label{eps}
 \end{align}
is needed. Using Eqs. (\ref{yy22}), (\ref{epsi}) and (\ref{eps}), we get another condition for Yukawa couplings:
\begin{align}
{\rm Im} [(\tilde{y}^\dag \tilde{y})_{32}^2 ] =1.7 \times 10^{-19} \frac{ {\rm keV}}{M_1} \frac{M_2}{10^9 \rm GeV}\frac{1}{g(M_3^2/M_2^2)}.
\label{Imyy}
 \end{align}
 
 \section{Constraints on parameters} \label{COP}
There are constraints on parameters $G_{FR}$, $M_I$, and $\Theta_1^2$ from our thermal history model and observations. We enumerate them as follows: 
\begin{enumerate}
\item $N_2$ must decouple while it is relativistic. From (\ref{Tf}) and $T_f>M_2$, we derive
 \begin{align}
 G_{FR}&\lesssim 10^{-23}{\rm GeV}^{-2}\left(\frac{10^9\rm GeV}{M_2}\right)^{3/2}. \label{GFR1}
  \end{align} 
 \item $N_2\to L_L \Phi $ must be a dominant decay channel of $N_2$. The decay diagram of Fig.\ref{NllN}(b) is similar to the decay of muon ($\mu\to e \overline{\nu_e}\nu_\mu$), so its rate can be estimated as
  \begin{align}
\Gamma_{N_2\to l_R \overline{l_R}\nu_R}&\sim 2\times\frac{G_{FR}^2}{192\pi^3}M_2^5
 \end{align}
(there is a factor 2 because $N_2$ is Majorana particle). Applying $\Gamma_{N_2\to l_R \overline{l_R}\nu_R}<\Gamma_{N_2}$, we obtain
 \begin{align}
 G_{FR}\lesssim 10^{-23}{\rm GeV}^{-2}\left(\frac{\rm keV}{M_1}\right)^{2}\left(\frac{10^9\rm GeV}{M_2}\right)^{3}. \label{GFR2}
  \end{align}
 This bound is similar to Eq. (\ref{GFR1}).
  
\item $N_2$ must decay when sphaleron transitions are still active \cite{Kuzmin:1985mm}. The temperature after $N_2$ decay is, using the decay rate (\ref{GN2}),
 \begin{align}
 T_{decay}&\sim \left(\frac{3}{8\pi g_*}\right)^{1/4}\sqrt{\Gamma_{N_2} M_{Pl}}\\
 &=4\times 10^{5}{\rm GeV}\frac{M_2}{10^9\rm GeV}\frac{\rm keV}{M_1}.
  \end{align}
  By $T_{decay}>100$GeV, we get 
  \begin{align}
  M_2\gtrsim 10^5{\rm GeV}\frac{M_1}{\rm keV}. \label{13Bez1}
   \end{align}
 As we will see later, a condition for sufficient lepton asymmetry production naturally satisfies this bound.
  
\item By phase space analysis of dwarf spheroidal galaxies and considering Pauli blocking, fermion dark matter must satisfy $M_1 \gtrsim 1$keV \cite{Gorbunov:2008ka}.
 
\item By comparing the Lyman alpha forest data and numerical simulations, the upper bounds on the initial average velocity of warm dark matter are derived \cite{Boyarsky:2008xj}. Those bounds can be converted to lower bounds of dark matter mass depending on their production mechanism. For thermal relic dark matter,  $M_1>1.5$keV.
 
\item $N_1$ is produced as a thermal relic, not through oscillations. The condition for the active-sterile mixing angle can be approximated as \cite{Asaka:2006nq}
 \begin{align}
 \sin^22\Theta_1 \lesssim 8\times10^{-8}\left(\frac{M_1}{\rm keV}\right)^{-2}. \label{Asa7.1}
  \end{align}
\item $N_1$ can decay into a photon and a light neutrino. Here we assume the radiative decay rate is same as the case in which the gauge interaction is not extended\footnote{In the left-right symmetric model, this assumption corresponds to taking the limit $|v_1v_2/v_R^2|\ll 1$ \cite{Bezrukov:2009th}. Reasonable parameters satisfy this inequality by the constraints (\ref{GFR1}) and (\ref{GFR2}).} \cite{Pal:1981rm},
\begin{align}
\Gamma_{N_1\to \nu \gamma}=\frac{9G_F^2\alpha}{256\pi^4}\Theta_1^2M_1^5.
 \end{align}
 Nondetection of those photons by x-ray observations gives a bound on $\Theta_1^2$  \cite{Abazajian:2001vt,Boyarsky:2005us,Boyarsky:2007ay,RiemerSorensen:2009jp} (see \cite{Drewes:2013gca} for a recent review). It can be roughly approximated as
 \begin{align} 
 \sin^22\Theta_1 \lesssim 5\times10^{-7}\left(\frac{M_1}{\rm keV}\right)^{-3.9}.
 \end{align}
\end{enumerate}

 \section{Combined analysis} \label{Ana}
In this section, we find parameters that satisfy all conditions described above, and compare them with observations. It is convenient to parametrize Yukawa couplings as 
  \begin{align}
 \tilde{y}v=iV_\nu^* (X_\nu^{\rm diag})^{1/2}R(M_R^{\rm diag})^{1/2},
 \end{align}
 where $X_\nu\equiv m_\nu-M_L,\; X_\nu^{\rm diag}\equiv V_\nu^T X_\nu V_\nu$ \cite{Akhmedov:2008tb}. The seesaw relation (\ref{Bezrukov46}) is transformed into an orthogonal condition for $R$,
 \begin{align} 
 RR^T=1, \label{RR}
 \end{align}
which is much easier to treat.  We rewrite the conditions for Yukawa couplings  (\ref{yy22}),  (\ref{Asa7.1}), and (\ref{Imyy}) by $X_\nu$ and $R$.  The results are
 \begin{align}
\sum_jX_j|R_{j2}|^2&=3.3\times 10^{-10}{\rm eV} \left( \frac{{\rm keV}}{M_1} \right)^2 ,
 \label{XR2}\\
\sum_j X_j|R_{j1}|^2 &<2\times10^{-5}{\rm eV}\frac{\rm keV}{M_1} ,\label{XR1}\\
{\rm Im}\left[\left(\textstyle \sum_j X_jR_{j3}^*R_{j2}\right)^2\right] &=1.5\times10^{-10}{\rm eV^2}\frac{ {\rm keV}}{M_1}\frac{10^9\rm GeV}{M_3}\frac{1}{g(M_3^2/M_2^2)}. \label{XR3}
 \end{align}
 
 Let us consider a simple case, $M_L=0$ first. We show this case does not give the correct neutrino mass matrix. From the orthogonality condition, at least one component of each column of $R$ must have an absolute value larger than $1/\sqrt{3}$. Considering (\ref{XR2}), one of the eigenvalues of $X_\nu=m_\nu$ must be equal or smaller than $\sim10^{-10}$eV. This is much smaller than observed values of neutrino mass differences, so masses of light neutrinos are determined except hierarchy. For a normal hierarchy case, using the experimental data \cite{Capozzi:2013csa}
 \begin{align}
 m_2^2-m_1^2=7.54\times 10^{-5}{\rm eV^2},\quad m_3^2-\frac{m_2^2+m_1^2}{2}=2.44\times 10^{-3}{\rm eV^2\quad (best\; fit)},
  \end{align}
  we obtain the masses of neutrinos
 \begin{align} 
 m_1\lesssim 10^{-10}{\rm eV}\ll m_2=8.7\times 10^{-3}{\rm eV},\; m_3=5.0\times 10^{-2}\rm eV.
 \end{align}
To satisfy Eqs. (\ref{XR2}) and (\ref{XR1}), the first and second columns of $R$ must be almost (1,0,0)$^T$, which contradicts the orthogonality of $R$. The $M_L=0$ case does not give the correct neutrino mass matrix. In short, two eigenvalues of $X_\nu$ must be much smaller than the orders of observed mass differences,
 \begin{align}
  X_1&\leq 3.3\times 10^{-10}{\rm eV} \left( \frac{{\rm keV}}{M_1} \right)^2, \\
 X_2&<2\times10^{-5}{\rm eV}\frac{\rm keV}{M_1}.
  \end{align}
 These conditions require $M_L$ to be nonzero.
 
 For concreteness, let us assume
 \begin{align}
X_1\ll X_2\ll X_3. \label{XXX}
 \end{align}
 Then, $R$ is almost determined as
  \begin{align}
 R&\simeq \begin{pmatrix} -R_{22}&1&-R_{32} \\1 &R_{22}&-R_{31}\\R_{31}&R_{32}&1  \end{pmatrix},
 \end{align}
 \begin{align}
 |R_{31}|^2&<2\times10^{-5}\frac{\rm eV}{X_3}\frac{\rm keV}{M_1} ,\\
 |R_{22}|^2&\leq 3.3\times 10^{-5}\frac{10^{-5}\rm eV}{X_2} \left( \frac{{\rm keV}}{M_1} \right)^2, \\
|R_{32}|^2&\leq 3.3\times 10^{-10}\frac{{\rm eV}}{X_3} \left( \frac{{\rm keV}}{M_1} \right)^2 ,\label{R32}\\
{\rm Im}\left[R_{32}^2\right] &=1.5\times10^{-10}\left(\frac{\rm eV}{X_3}\right)^2\frac{ {\rm keV}}{M_1}\frac{10^9\rm GeV}{M_3}\frac{1}{g(M_3^2/M_2^2)}. \label{ImR}
  \end{align} 
 $R_{22}$, $R_{31}$, and $R_{32}$ are much smaller than 1, so $R$ is approximately orthogonal.  From (\ref{R32}), (\ref{ImR}), and using $|{\rm Im}[R_{32}^2]|\leq |R_{32}|^2$, we obtain
 \begin{align}
 X_3&\geq 0.45{\rm eV}\frac{M_1}{ {\rm keV}}\frac{10^9\rm GeV}{M_3}\frac{1}{|g(M_3^2/M_2^2)|} .\label{X3}
  \end{align}
 This partly justifies the assumption (\ref{XXX}). If there is no tuning between the two terms in the right-hand side of (\ref{Bezrukov46}), $X_{1,2,3}$ is smaller than $O(1)$ eV. Assuming also $M_3>2M_2$, $g(M_3^2/M^2_2)$ can be approximated as
 \begin{align}
 g(M_3^2/M^2_2)\sim -\frac{3M_2}{2M_3}.
  \end{align}
  From Eq. (\ref{X3}) we obtain
 \begin{align}
 M_2 \gtrsim O(10^{8})\rm GeV.
  \end{align} 
  This is the reason for dividing $M_2$ by 10$^9$GeV. The condition from the sphaleron freeze-out temperature (\ref{13Bez1}) is satisfied naturally.
 
\begin{figure}[t]
  \centering
 \includegraphics[width=13cm]{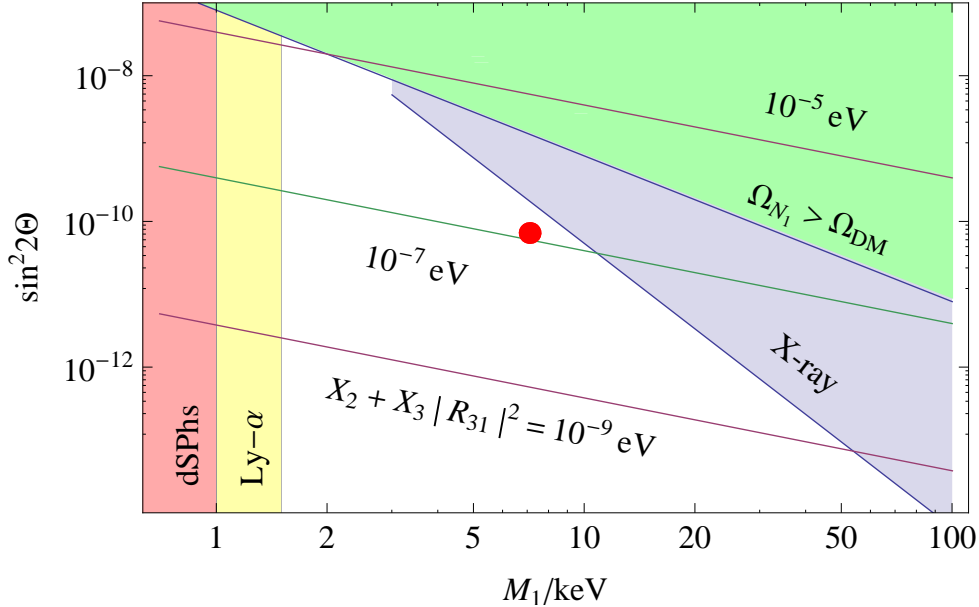}
  \caption{Active-sterile mixing angle of $N_1$. We plotted our result for $X_2+X_3|R_{31}|^2$=10$^{-5}$eV, 10$^{-7}$eV, and 10$^{-9}$eV. The colored region is excluded by observation (see Sec. \ref{COP}). The circle at $(M_1,\;\sin^22\Theta_1)=(7{\rm keV}, 7\times 10^{-11}) $ shows the possible detection reported in \cite{Bulbul:2014sua,Boyarsky:2014jta}.}  \label{Res}
\end{figure}

 We compare the results with observations, especially x-ray observation. The mixing of $N_1$ is 
  \begin{align}
\sin^22\Theta_1=4\frac{X_2+X_3|R_{31}|^2}{M_1}.
 \end{align}
 Note that $X_1|R_{11}|^2\ll X_2, X_3|R_{31}|^2$ so $X_1$ does not contribute to $\Theta_1^2$. We have plotted this result in Fig. \ref{Res} with various constraints. If $X_2+X_3|R_{31}|^2$ is smaller than 10$^{-9}$eV, the dark matter mass can be heavier than 50 keV. This is different from the $\nu$MSM,  
which needs $M_1\lesssim 50$ keV due to a constraint from maximal lepton asymmetry \cite{Laine:2008pg,Canetti:2012kh}.
 
 Recently, a possible sterile neutrino decay signal was found in x-ray multi mirror mission data \cite{Bulbul:2014sua,Boyarsky:2014jta}. These papers report
 \begin{align}
 M_1&\simeq7\rm keV,\\
 \sin^22\Theta_1&\simeq7\times 10^{-11}.
  \end{align}
  These values correspond to
  \begin{align}
  X_2+X_3|R_{31}|^2\simeq1\times 10^{-7}\rm eV
   \end{align}
   in our model. This point is shown in Fig. \ref{Res}. The much better energy resolution of the satellite ASTRO-H \cite{Takahashi:2012jn} will be able to distinguish the decaying dark matter line from plasma emission lines.
\section{Conclusion}
\label{sec:conc}
We considered a model in which heavy neutrinos $N_1$ and $N_2$ are in thermal equilibrium due to a new gauge interaction at the temperature $T\gtrsim 10^{10}$ GeV. In this case, dark matter $N_1$ is overproduced, so we supposed $N_1$ was diluted by out-of-equilibrium decay of $N_2$. This decay also produces the observed matter-antimatter asymmetry. Because of the condition from entropy production, and an oscillation constraint of $N_1$, the two eigenvalues of $X_\nu\equiv m_\nu-M_L$ have to be much smaller than the observed mass differences. This means $M_L$ is needed in our scenario. We determined the possible range of eigenvalues of $X_\nu$, masses of right-handed neutrinos $M_I$, and parameter of Yukawa couplings $R$, which can simultaneously explain three beyond-the-SM phenomena: neutrino masses, the BAU, and dark matter. In this model, a wider range of dark matter mass $M_1$ is allowed than in the DW mechanism and the $\nu$MSM. Our model can explain the recent anomaly in x-ray observation by taking $M_1\simeq 7 {\rm keV},\; X_2+X_3|R_{31}|^2\sim1\times 10^{-7}\rm eV$. In the near future, the ASTRO-H mission will clarify whether this signal is from a dark matter decay or not.


\section*{Acknowledgments}

The author is grateful to Masahiro Kawasaki, Takehiko Asaka, and Euan Richard for useful discussions and comments.


\bibliographystyle{JHEP}
\bibliography{/Users/t/Documents/Research/ref}

\end{document}